\documentclass[12pt,preprint]{aastex}
\shorttitle{NGC 6543 revisited}
\shortauthors{Luridiana & P\'erez}
\def\greatsim{\mathrel{\hbox{\rlap{\hbox{\lower4pt\hbox{$\sim$}}}\hbox{$>$}}}}

\begin{document}
\title{Physical conditions in the O$^{++}$ zone from ISO\footnote
{Based on observations with ISO, an ESA project with instrument funded 
by ESA Member States and with the participation of ISAS and NASA.} 
\ and HST data. NGC 6543 revisited.}

\author{V. Luridiana and E. P\'erez}
\affil{Instituto de Astrof\'\i sica de Andaluc\'\i a (CSIC), Granada, Spain}
\author{M. Cervi\~no}
\affil{Instituto de Astrof\'\i sica de Andaluc\'\i a (CSIC), Granada, Spain}
\affil{Laboratorio de Astrof\'\i sica Espacial y F\'\i sica Fundamental (INTA),
Madrid, Spain}

\begin{abstract}

We revise the physical conditions in the O$^{++}$ zone 
of the planetary nebula NGC 6543,
obtaining two different estimates of the electron temperature ($T_e$)
and one estimate of the electron density ($N_e$).
The electron temperature is computed by means of two independent methods, 
the nebular-to-auroral ratio [O{\sc~iii}] $\lambda\,5007$/$\lambda\,4363$, 
and the diagnostic diagram that combines $\lambda\,5007$ to 
the [O{\sc~iii}] infrared lines $52\,\mu$ and $88\,\mu$. 
The optical and infrared fluxes have been obtained 
from archive HST/WFPC2 images and ISO LWS spectra respectively,
and the continuum intensity in the optical
has been measured from narrow-slit spectra obtained with the Isaac Newton Telescope
at La Palma.
The measured continuum intensity is 
higher than predicted by recombination theory
under the hypothesis that all the Ly$\alpha$ photons either 
escape or are destroyed.
This fact can be explained in terms of 
an enhancement of the 2-photon continuum due to Ly-$\alpha$ conversion,
a process that depends strongly on the local structure of the nebula.
Alternative possibilities, outside the framework of
recombination theory, have also been considered:
e.g., the optical tail in the X-ray emission of a very hot plasma,
and dust scattering of stellar radiation,
but these hypotheses are not supported by quantitative estimates.

While the electron temperature and density derived from the diagnostic diagram
agree with the most recent determination,
the temperature derived from $\lambda\,5007$/$\lambda\,4363$, 
$T_e($O$^{++})_{opt}$, is somewhat smaller than previously published values.
We discuss several technical issues that contribute to the overall
uncertainty in our results, focusing on the instrumental effects 
that might bias the [O{\sc~iii}] $\lambda\,4363$ intensity. 
We also discuss the effects of the collisional de-excitation 
of the O$^{++}$ ground terms on the relation between $T_e($O$^{++})_{opt}$ 
and $T_e($O$^{++})_{IR}$.

\end{abstract}


\hfill{\today}

\section{Introduction}

Knowledge of physical conditions ($T_e$ and $N_e$) is
an essential ingredient in the understanding of planetary nebulae.
Information on the structure, chemistry and kinematics 
of these objects comes from individual diagnostics
and from the cross-comparison of different diagnostics.
In this work, two different diagnostics will be used to 
investigate the physical conditions
in the planetary nebula NGC 6543:
the temperature-sensitive 
[O{\sc~iii}] ratio $I(\lambda\,5007)/I(\lambda\,4363)$,
and the ($T_e$, $N_e$) diagnostic diagram
based on the [O{\sc~iii}] optical and infrared lines, 
which was first proposed by \citet{DLW85} \citep[henceforth DLW; see also][]{D83}.

The first method, the one based on the 
$I(\lambda\,5007)/I(\lambda\,4363)$ ratio, 
is one of the most popular ones.
Advantages of this method are that the lines involved are bright,
and close enough to be easily observed with the same instrumental setup;
a caution to be taken is that this ratio selectively samples
the hottest zones, so that in presence of strong
temperature fluctuations the derived temperature, $T_e($O$^{++})_{opt}$, 
is biased towards the highest values \citep[see, e.g.,][]{P95}.
Furthermore, this ratio measures the temperature of the O$^{++}$ region,
which is not necessarily representative of the whole observed volume.
The second method, the diagnostic diagram by DLW, allows the simultaneous
determination of the electron density and 
temperature by combining the optical and the FIR fine-structure [O{\sc~iii}] lines.
The two infrared lines [O{\sc~iii}] 52$\mu$ and 88$\mu$
arise from low-energy levels of the ground term, so their ratio 52$\mu$/88$\mu$
is almost insensitive to the temperature, 
while the critical densities of their upper levels are different,
so the ratio is strongly sensitive to density.
On the other hand, a ratio like $\lambda\,5007$/52$\mu$ is strongly
dependent on temperature, due to the very different excitation
energies of the upper levels of the two lines,
and it has a weak dependence on density too,
due to the difference in the critical densities. By combining the two 
line ratios, one can obtain a simultaneous measurement of the electron temperature, 
$T_e($O$^{++})_{IR}$, 
and density, $N_e($O$^{++})_{IR}$ (we will use throughout 
the text the subindex {\it IR} when referring to these $T_e$ and $N_e$ determinations,
though an optical line is involved also;
this choice responds to simplicity criteria, and highlights the
fact that the density is, to a good approximation,
a function of the ratio of the two infrared lines alone).
One of the main advantages of this method is
its internal consistency, since all the lines involved belong to the same ion,
so that the same volume is mapped to derive both temperature and density
(although the optical and infrared emissivities might follow different
distributions within the O$^{++}$ zone
if temperature or density fluctuations are present).
Two major disadvantages are consequences of the lines involved lying far apart in the spectrum:
first, the lines must be observed with different instruments, 
a procedure implying the risk of 
systematic offsets in the calibration;
and second, any uncertainty in the reddening correction is amplified by the 
large wavelength baseline.

By means of these methods, we want to reassess the physical conditions in 
the bright core of NGC 6543.
A similar analysis was performed on a small sample of PNe,
including NGC 6543 itself, by DLW.
These authors used infrared data obtained with the Kuiper Airborne Observatory
(KAO), and complemented them with optical spectra from the Lick Observatory
Crossley reflector;
the analysis was later repeated by \citet{Dal95} (henceforth DHEW),
with quite different results.
In the work we present here, the intensity of the infrared lines is obtained from ISO archive spectra,
and the intensity of [O{\sc~iii}] $\lambda\,5007$ and
[O{\sc~iii}] $\lambda\,4363$ from HST archive images.
Additionally, H$\alpha$ and H$\beta$ HST images 
were used to determine the reddening correction to be applied to the other images,
and narrow band HST images around $\lambda\, 6584$ were used to correct the H$\alpha$ images for the 
[N{\sc~ii}] contribution.
Finally, long-slit, ground-based data obtained with the 
Isaac Newton Telescope at the Observatorio del Roque de los
Muchachos
were used to correct the HST images 
for the contribution of other lines and the continuum.
These data will be described in Sections~\ref{sec:ISOdata},~\ref{sec:INTdata} and~\ref{sec:HSTdata}.

\subsection{NGC 6543}

NGC 6543 (the `Cat's eye') is a very well known, low-excitation planetary nebula.
Many studies have been devoted to understand its complex structure;
an incomplete list of the most recent ones includes \citet{Ral99,BWH01,Hal01}.
The bright core of the nebula can be roughly described
as the superposition of two ellipses, equal in shape but perpendicular to each other
\citep{MS92}.
This region is about $25''$ in diameter, and can be seen in direct images in
the Balmer and [O{\sc~iii}] lines. 
Other structures form part of the nebula:
surrounding the bright core, a series of dim concentric rings is visible
in [O{\sc~iii}], extending at least out to $70''$ \citep{BWH01};
in turn, the rings are surrounded by a giant filamentary halo of about $300''$ in diameter
\citep{BWH01,MCW89};
X-ray emission coming from the bubble surrounding
the central star has been recently reported by \citet{Cal01};
the variability of the central star has been investigated by \citet{BWH01}.

\section{The ISO data\label{sec:ISOdata}}

NGC 6543 was routinely observed by ISO for calibration purposes,
thereby a large set of spectra is available.
We selected 20 out of the 92 Long-Wavelength Spectrometer (LWS) 
spectra of NGC 6543 available in the ISO data archive, all of them in the AOT L01 observing mode; 
the selected spectra were taken during two separate periods, 
close to the beginning and the end of ISO operational life respectively
(see Table~\ref{tab:ISOjournal}). 
A detailed explanation of the LWS structure and data sets is provided
in {\em The ISO Handbook, vol. IV, LWS - The Long Wavelength Spectrometer}
\citep{Gry01}.

Each spectrum is comprised of 10 independent wavelength ranges, 
which correspond to the 10 detectors that form the LWS. 
These wavelength ranges overlap partially. Both emission
lines considered by us fall in the overlap region of two consecutive ranges: 
i.e., [{\sc O~iii}] 52$\mu$ falls on both
the SW1 and SW2 detectors, while [{\sc O iii}] 88$\mu$ falls
on both the SW5 and LW1 detectors.
Additionally, the spectrum of each of these 10 wavelength ranges is 
composed of 6 sub-exposures, that correspond to independent scans of the detector.
Thus, we can in principle perform, in our sample, (20 spectra) 
$\times$ (2 detectors) $\times$ (6 scans) = 240
independent measures for any of the two emission lines of interest,
or 120 with each detector.
In practice,
the number of independent measurements of a given line with a given
detector is smaller than 120,
due to the following reasons:
in a few cases, one or two out of the six scans of a given exposure 
do not span the whole detector range, failing to include the line;
in other cases the quality of the spectrum is very poor,
so that the line-intensity measurement is unfeasible. 
However, the measurements are still numerous enough 
(77 on average per line and detector)
to adequately sample the data distribution.

\begin{figure}
\epsscale{0.8}
\plotone{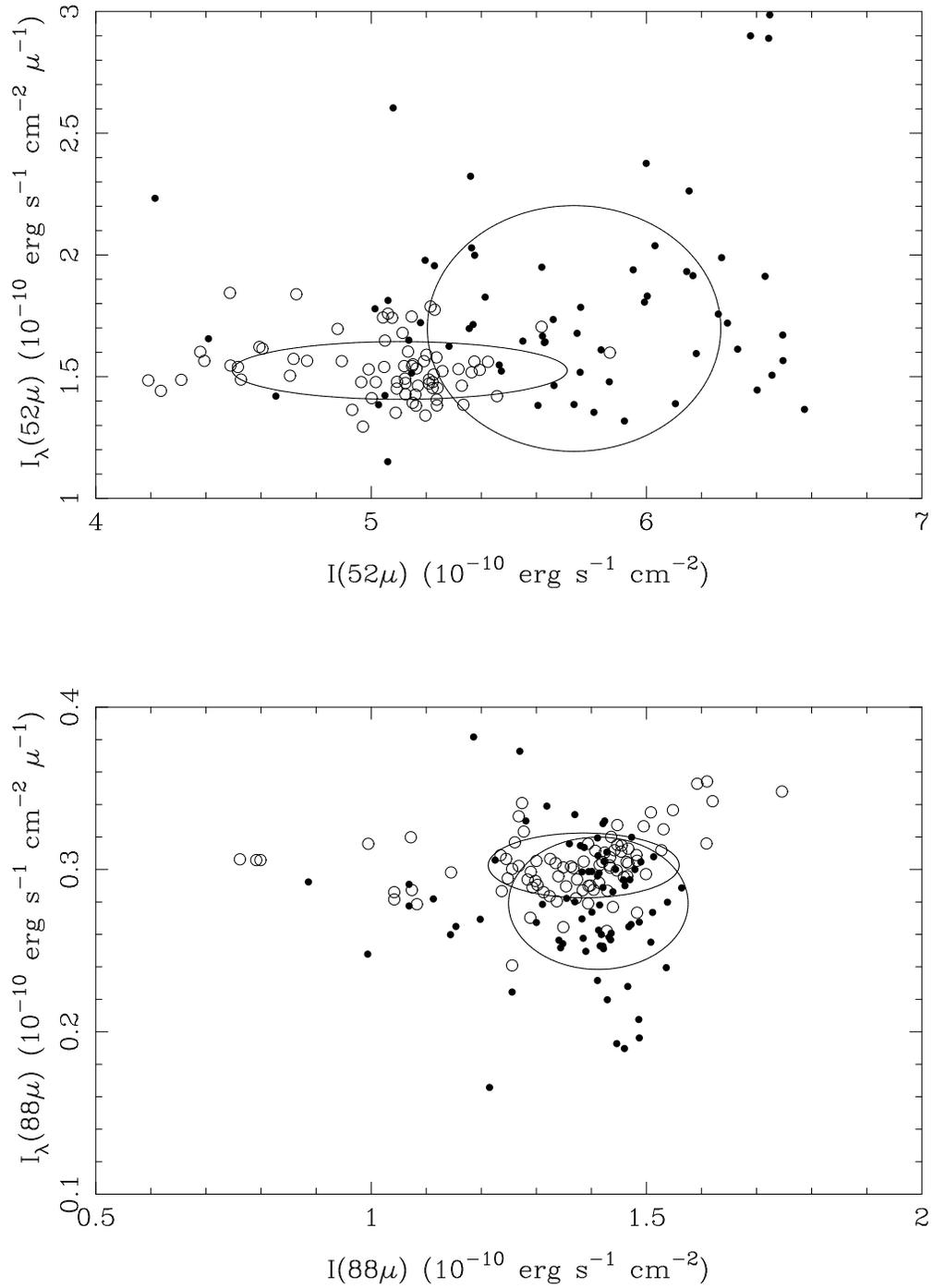}
\caption{ISO 52$\mu$ and 88$\mu$ line vs. continuum intensities for our sample,
with the 1-$\sigma$ ellipses overplotted. 
Upper panel: dots are SW1 data, open circles SW2 data. 
Lower panel: dots are SW5 data, open circles LW1 data. \label{fig:iso}}
\end{figure}

The data were reduced using the latest version of the 
ISO automatic analysis pipeline available in November 2001.
The flux measurements were performed by model gaussian fitting.
Figure~\ref{fig:iso} represents the continuum flux versus emission line
intensity for all the measurements of the two features,
52$\mu$ and 88$\mu$.
In the upper plot, the filled dots and the open circles 
represent the data taken with the SW1 and the SW2 detectors 
respectively; in the lower plot, the dots and the open circles 
represent the data taken with the SW5 and the LW1 detectors 
respectively.
The ellipses delimit the region within 1-$\sigma$ from the median values.

Table~\ref{tab:ISOdata} summarizes the median and dispersion values
of the continuum and line measurements for the four detectors.
The data of each epoch are shown both as individual data sets, 
and combined together.
The following conclusions can be drawn from the analysis of the data sample:

\begin{itemize}
\item[a)] the SW1 and SW5 continuum data  are much more spread out than 
   the corresponding SW2 and LW1 data;
\item[b)] the median intensity values of the 88$\mu$ line obtained
   with SW5 and LW1 are in excellent agreement, 
   whereas the median intensity values of the 52$\mu$ line 
   obtained with SW1 and SW2 are more discrepant,
   although they still agree at the 1-$\sigma$ level;
\item[c)] the median intensity values of the 52$\mu$ line obtained with SW1 
   suffered a drift of more than 1$\sigma$ from Epoch I (June 1996 + July 1996) 
   to Epoch II (December 1997 + January 1998). 
   This long term drift is known and documented in the ISO LWS Handbook. 
\end{itemize}

Based on these results,
we decided to perform our analysis on the line 
intensities measured with the SW2 and LW1 detectors only, 
neglecting the data obtained with the other two detectors.
The exclusion of SW1 is primarily motivated by the drift in the
line flux data, which make them unreliable.
On the other hand, the exclusion of SW5 is only motivated
by the larger spread in the continuum data
with respect to those of the LW1 detector,
a feature that does not necessarily imply a lower quality
of the line fluxes obtained with this detector.
In fact, since SW5 has a spectral resolution twice that of LW1, 
the recorded dust continuum level per resolution element is lower,
making this detector more sensitive to the dark current subtraction:
as a result, a fraction of the larger spread in the SW5 continuum data may
be caused by the uncertainties in the subtraction of the dark current,
an uncertainty not affecting the line fluxes.
Nonetheless, we preferred to adopt a conservative approach and exclude
both SW1 and SW5 from our analysis.
With such a choice, the adopted line intensities are:

\begin{equation}
I(52\mu)=(5.10\pm 0.61)\times 10^{-10}\ {\rm erg\,sec}^{-1} {\rm cm}^{-2}
\end{equation}
and 
\begin{equation}
I(88\mu)=(1.39\pm 0.17)\times 10^{-10}\ {\rm erg\,sec}^{-1} {\rm cm}^{-2},
\end{equation}
where the quoted value is the median of the sample, 
and the uncertainty is the dispersion in the data,
which does not include possible systematic error sources, such as, e.g.,
the flux calibration.
These values are in excellent agreement with those reported by \citet{Lal01},
$I(52\mu)=(5.11\pm 0.13)\times 10^{-10}$ erg sec$^{-1}$ cm$^{-2}$ 
and 
$I(88\mu)=(1.49\pm 0.04)\times 10^{-10}$ erg sec$^{-1}$ cm$^{-2}$, 
although \citet{Lal01} quote rather small uncertainties.

A flux correction factor should be applied to take into account the extension of the source,
since the ISO LWS analysis pipeline is calibrated on Uranus and it 
assumes that the target is a point source.
This flux correction factor accounts for the relation between
the ISO telescope point spread function and the LWS instrumental profile. 
DLW applied the corresponding flux correction to their KAO data,
a correction that was quite small, amounting to only 4\% for the $52\,\mu$ line. 
In our case, the fraction of flux diffracted out of the aperture
is even smaller, and it is therefore 
negligible with respect to the other uncertainties involved in the process,
such as, e.g., the flux calibration itself.
Consequently, we did not apply any correction for this effect.

\section{The Isaac Newton Telescope long-slit data\label{sec:INTdata}}

NGC 6543 was observed spectroscopically as part of a wider study of PNe
on 1995 July 8 and 9. 
We used the Intermediate Dispersion Spectrograph
attached to the
2.5m Isaac Newton Telescope (INT) at the Observatorio del Roque de los
Muchachos,
on La Palma. The $4' \times 1 ''$ slit was placed at position
angles 5$^{\circ}$ and 20$^{\circ}$,
and included the central star. 
The R1200B grating and the TEK3 CCD
in the 235 mm
camera provide a dispersion of 0.85 \AA/pixel and a spatial sampling of
0.71$''$/pixel.
The seeing was stable at 1.1$''$, and the integration time was 1000
s.
The spectra were calibrated using the standard steps of flat fielding,
bias subtraction, wavelength calibration, and flux calibration using
flux
spectroscopic standard stars observed during the same night and 
with the same setup,
but
through a 10 arcsec wide slit;
the spectra were also corrected for light dispersion inside the spectrograph.
The inset in the top panel of
Figure~\ref{fig:spec}
illustrates the very high quality of these data, showing 
a logarithmic plot
of a sample nebular spectrum in the wavelength range 3400--4500 \AA.
These data form part of a much larger data set that will be presented
elsewhere.

\section{The HST images\label{sec:HSTdata}}

The HST data archive at ST-ECF was queried for WFPC2 images of NGC 6543. 
We retrieved images through the narrow band filters F437N, F487N, F502N, 
F656N, and F658N,
centered on the emission lines [O{\sc~iii}] $\lambda\,4363$, H$\beta$,
[O{\sc~iii}] $\lambda\,5007$, 
H$\alpha$, and [N{\sc~ii}] $\lambda\,6584$ respectively. 
Table~\ref{tab:HSTjournal}
summarizes these data sets. For each filter, the different exposures were 
combined with the IRAF\footnote{IRAF is distributed by the National Optical 
Astronomy Observatories, which are operated by the Association of Universities 
for Research in Astronomy, Inc., under cooperative agreement with the National
Science Foundation.}/STSDAS command {\tt crrej}, to obtain a final image 
at each waveband free of cosmic rays.
In each of these images, the fluxes were measured within the PC frame
of the WFPC2.
Similar analyses have been performed by \citet{LHB97} and \citet{Hal01}
for NGC 6543, and \citet{Ral02} for NGC 7009.
Since our results differ from those of both \citet{LHB97} and \citet{Hal01},
we will describe our procedure in full detail, 
so as to allow future identification of the causes behind the discrepancies.
The calibration is conceptually straightforward, but the procedure
contains several potential pitfalls, such as, e.g.,
the zeropoint calibration, the adoption of the proper width for the filter,
or the estimation of the contribution to the flux of the continuum and of neighbouring lines.

\subsection{Photometric zeropoint calibration}

The zeropoint of an instrument is defined as the
intrinsic flux of an object producing one count per second.
The raw data number (DN: the ratio between the electron count
in each pixel, and the gain factor, which is 7 in our case)
obtained from the HST images must
be multiplied by the appropriate zeropoint in order
to obtain the flux in energy units.
The zeropoint for the WFPC2 is codified in the header word 
PHOTFLAM. As many other calibration parameters, PHOTFLAM
has been periodically modified, as the calibration
routine improved.
The PHOTFLAM values can be found either in the header of the images, 
or in the HST-WFPC2 Data Handbook \citep{WDH02}.
Unfortunately, the two sources give different values;
the difference is quite sizeable (almost 4\%) in the case of
the H$\alpha$ filter F656N, and smaller than 1\%
for the other filters. 
As the HST helpdesk advised us to do, we sticked to the PHOTFLAM values of
the WFPC2 Data Handbook, which are listed in Table~\ref{tab:HSTjournal}.

\subsection{Width of the filter\label{sec:width}}

The narrow WFPC2 filters are characterized by several parameters.
One of them is the filter's rectangular width RECTW,
defined as the width in \AA\ of a rectangle with the same total area
as the total transmission curve, which describes the transmittance 
of the system, and height 
given by the peak in the curve.
Since the DN is given per unit wavelength,
the data measured in the WFPC2 images must be multiplied by
the RECTW value of the corresponding filter.
As already noted by \citet{Ral02}, RECTW should not be confused with
the $\delta\lambda$ values given in the WFPC2 Instrument Handbook \citep[][Appendix 1]{Bal01}.
RECTW can be computed for each filter with the IRAF task {\tt synphot/bandpas}.
The values for the filters considered in this work
are given in Table~\ref{tab:HSTjournal}. 

\subsection{Field of view}

The PC frame is a square of 800 $\times$ 800 pixels,
corresponding to about $37''\times37''$.
The central $40 \times 38$ pixels were excluded,
to avoid contamination by the central star.
The outermost pixel rows and columns are affected by optical problems 
related to the spherical aberration of the telescope.
Due to the low photometric quality of these pixel rows
they were excluded from our analysis,
and the flux was measured in a square of 690 $\times$ 670 pixels,
corresponding to about $32''\times30''$.
This square is smaller than the field of view of the ISO LWS,
which has a diameter of about $80''$ at the wavelengths of interest.
However, the flux emitted by the nebula is completely dominated by 
the innermost $30''$, which accounts for more than 97 percent of the total 
(Figure~\ref{fig:OIIIradi}), 
and the two instruments effectively sample the same area
\citep[this is also shown, e.g., by the brightness distribution of
the [Ne{\sc~iii}{]} line at $15\,\mu$: see][]{Pal99}.
This issue is also addressed by DLW,
who match the KAO $40-50''$ aperture by
taking the ground based observations of [O{\sc~iii}]$\lambda\,5007$ and H$\beta$ 
through $45''$ and $70''$ diameter apertures, and using the [O{\sc~iii}]$\lambda\,4363$
fluxes by \citet{BDO74} through a similarly wide aperture.

\subsection{Flux contribution of neighbouring lines\label{sec:linecon}}

A fraction of the flux measured in each filter is 
contributed by lines other than the principal line.
In the case of the four filters we considered, 
F487N and F502N contain only H$\beta$ and $\lambda\,5007$ respectively,
whereas F437N includes H$\gamma$ $\lambda\,4340$ and He{\sc~i} $\lambda\, 4388$,
and F656N includes [N{\sc~ii}] $\lambda\lambda\, 6548$, 6584.
In this section, we describe the procedure followed
to subtract these contributions from the total measured fluxes.

The total flux of [N{\sc~ii}] $\lambda\,6584$ was computed from
an image taken with the F658N filter,
and then subtracted from the F656N image taking into account
the transmission efficiency of the filter at the position
of $\lambda\,6584$.
The [N{\sc~ii}] $\lambda\,6548$ contribution was computed theoretically
from the flux of [N{\sc~ii}] $\lambda\,6584$.
Since the F658N filter is, in turn, contaminated by H$\alpha$,
the decontamination procedure was iterated 
to give decontaminated H$\alpha$ and [N{\sc~ii}] $\lambda\,6584$ flux values.
After this procedure was applied, 
the total H$\alpha$ flux lowered by about 3\%.

The total flux in the F437N filter includes
contributions of H$\gamma$ $\lambda\,4340$ and He{\sc~i} $\lambda\,4388$.
Both contributions are difficult to estimate,
since they depend on
the exact position of the line in the filter.
In the case of He{\sc~i} $\lambda\,4388$, the intensity of the line itself
is also unknown, whereas the total intensity of H$\gamma$ 
can be computed from $I($H$\alpha)$.
We estimated these contributions
multiplying the nebular spectrum of the long-slit
spectra (with the central star excluded)
by the F437N transmission curve, and then measuring separately each component
on the resulting pattern.
The top panel in Figure~\ref{fig:spec} shows the nebular spectrum of NGC 6543 
along position angle 5$^{\circ}$
in the wavelength region around H$\gamma$ 4340, [O{\sc~iii}] 4363 and He{\sc~i} 4388. 
The dashed line is the actual spectrum, while the full line is this spectrum
convolved with the HST F437N filter response (the total system peak response
of 0.03 \citep{Bal01} has been divided out so as to make the two spectra comparable). 
The inset shows a sample of the spectrum along the slit in a logarithmic scale,
to provide an indication of the high S/N of this data set.
We found that H$\gamma$ and He{\sc~i} $\lambda\,4388$ contribute on average
12\% and 4\% 
respectively to the total flux in F437N, while [O{\sc~iii}] $\lambda\,4363$ provides 
on average 26\% of the flux, 
with a typical deviation of 3\%.
This is plotted in the bottom panel of Figure~\ref{fig:spec}, which shows 
the contribution of $\lambda\,4363$ to
the total intensity in F437N, normalized to 100
(dotted circles, left handside scale; the central $\pm 3''$ are
excluded from the analysis);
the rest of the flux detected through the filter is contributed by H$\gamma$ and He{\sc~i} 4388,
other very faint emission lines, and by the underlying nebular 
continuum (see top panel). Also plotted for reference 
is the variation of the [O{\sc~iii}] 4363 line intensity along 
the slit (solid line, right hand side scale).

\subsection{Continuum subtraction}

The underlying continuum contributing within each of the HST images 
can be ascertained either theoretically or empirically.
The first method requires a detailed computation of the efficiency
of all the processes involved, while the second consists in measuring 
on a spectrum the relative importance of each line with respect to the 
underlying continuum,
with the filter's transmittance curve taken into account.
We chose to do it both ways, since the comparison between
the results illustrates
quantitatively the difficulties of an accurate theoretical computation
of the nebular processes.

\subsubsection{Theoretical continuum contribution}\label{sec:cont_theo}

The processes contributing to the nebular continuum flux are {\sc H~i}, {He\sc~i},
and {He\sc~ii} recombination, bremsstrahlung, and 2-photon decay.
The relative importance of these processes depends on both the wavelength and the
nebular electron temperature, and on the {He\sc~i} and {He\sc~ii} abundances.
Table~\ref{tab:continuum} shows the frequency dependence of the 
continuum-emission coefficient, $\gamma_\nu$, for each process 
at the three wavelengths of interest, in units 10$^{-40}$ erg cm$^3$ sec$^{-1}$ Hz$^{-1}$; 
the data have been obtained interpolating in the ($\nu$, log $\gamma_\nu$) 
plane between the values listed by \citet{BM70} at $T_e=8000$ K.

The 2-photon continuum intensity is quite difficult
to compute accurately, as it depends not only on the local physical
conditions $N_e$ and $T_e$, but also on the 
fate of the Ly$\alpha$ photons produced by recombination.
In optically thin or dusty nebulae, all the electrons
on the 2$^2$P level produce a Ly$\alpha$ photon
that eventually escapes from the nebula or is destroyed by dust,
whereas those
on the 2$^2$S level decay to the ground state producing
2-photon continuum,
the population on 2$^2$S and 2$^2$P levels being determined
by the recombination cascade and the mutual collisional transitions.
In practice, Ly$\alpha$ photons can be scattered repeatedly
before they escape from the nebula, and at each scattering
the absorbing electron has a finite probability of suffering a 2$^2$P $\rightarrow$ 2$^2$S
transition and decaying via 2-photon emission.
Thus, with given local physical conditions ($N_e$, $T_e$),
the actual 2-photon emission can vary from a minimum value
(corresponding to the case in which all the Ly$\alpha$ photons
are either destroyed by dust or escape from the nebula)
to a maximum value (corresponding to the case in which 
all the Ly$\alpha$ photons are eventually converted to
2-photon continuum).
In Table~\ref{tab:continuum} and Figure~\ref{fig:continuum}
we represent, as a function of the wavelength,
the two limiting values for the 2-photon continuum,
and the corresponding values for the total continuum.
The impact of the occurrence of Ly$\alpha$ conversion on the total
continuum level depends on the wavelength considered;
in the case of the wavelength range around
$\lambda$ 4363, where 2-photon emission is a dominant process,
Ly$\alpha$ conversion can enhance the total continuum by as much as a 
factor 2.5. 
This would be the asymptotic case of a very thick, dust-free nebula,
and represents just an upper limit to be considered in the following.
We assumed in the calculations $T_e=8000$ K and
$N_e=1650$ cm$^{-3}$; the density corresponds to the observed $52\mu/88\mu$ ratio:
see Section~\ref{sec:NT_IR}.
The collisional rates listed by \citet{O89}
for the 2$^2$S $\rightarrow$ 2$^2$P transition
have been linearly extrapolated in temperature down to $T_e=8000$ K;
this might seem a rough approximation, but our results
are fairly insensitive to it, both because the collisional data
at $T_e=10000$ and $20000$ K are quite similar, and because
collisions are outnumbered by spontaneous decays at these nebular
conditions anyway.

The {He\sc~ii} recombination continuum is not listed because it is not relevant
for NGC 6543, due to the low He$^{++}$ abundance 
\citep[$N({\rm He}^{++})/N({\rm H}^+)\sim 10^{-4}$,][]{Hal00}.
On the other hand, $N({\rm He}^{+})/N({\rm H}^+)\sim 0.11$,
so that the {He\sc~i} recombination continuum cannot be neglected.

The continuum-emission coefficient $j_\nu$, measured in 
erg sec$^{-1}$ cm$^{-3}$ Hz$^{-1}$ ster$^{-1}$,
is related to $\gamma_\nu$ through $j_\nu = N_+N_e\gamma_\nu/4\pi$.
Figure~\ref{fig:continuum} shows $j_\nu$ normalized to $N({\rm H}^+)N_e/4\pi$, 
i.e., the {He\sc~i} recombination contribution
has been multiplied by the factor $N({\rm H}^+)/N({\rm He}^{+})$,
so that the contributions of the three processes to the continuum
can be directly compared.
The specific intensity emitted through each process
depends on the width of the filter's passband RECTW:
$I_\nu=\int_{\rm RECTW} j_\nu d\nu = N_+N_e/4\pi\int_{\rm RECTW}\gamma_\nu d\nu$.

The continuum contribution is very important in filters centered on weak lines,
such as [{\sc O~iii}] $\lambda\,4363$,
but it barely affects the intensity in filters centered on
strong lines, such as H$\alpha$ or [{\sc O~iii}] $\lambda\,5007$.
Our computations show that, at $T_e=8000$ K
and neglecting Ly$\alpha$ conversion,
the continuum yields 40$\%$ of the total intensity 
in the F437N filter around [{\sc O~iii}] $\lambda\,4363$, 
0.3$\%$ of the total intensity in the F502N filter around [{\sc O~iii}] $\lambda\,5007$, 
and 0.5$\%$ of the total intensity in the F656N filter centered on H$\alpha$. 
As anticipated before, we also measured the continuum contribution
on long-slit spectra, and compared it to our theoretical
computations. Our findings are discussed in next section.

\subsubsection{Continuum contribution from direct measurement}\label{sec:cont_obs}

\begin{figure}
\epsscale{0.8}
\plotone{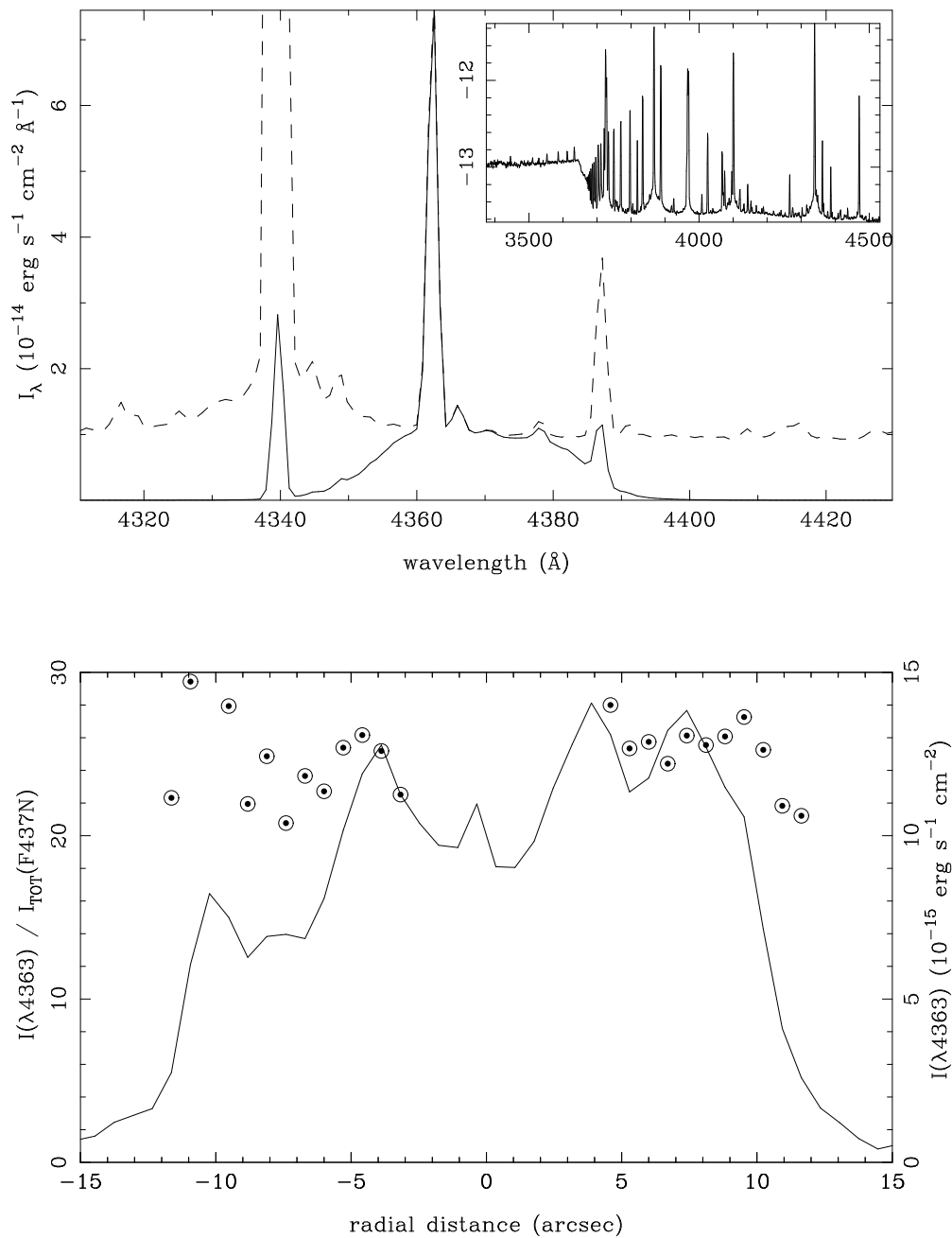}
\caption{(top) Nebular spectrum of NGC 6543 along position angle 5$^{\circ}$
in the wavelength region around $\lambda\,4363$.
The dashed line is the actual spectrum; the full line is the spectrum
convolved with the F437N filter response (the total system peak response
of 0.03 has been divided out).
The inset shows a sample of the spectrum along the slit in a logarithmic scale.
(bottom) The dotted circles (left handside scale) show the percent of 
the total intensity of radiation through the HST filter F437N that is
contributed by the $\lambda\,4363$ line intensity.
Also plotted is the variation of the $\lambda\,4363$ intensity along 
the slit (full line, right hand side scale).\label{fig:spec}}
\end{figure}

\begin{figure}
\epsscale{0.8}
\plotone{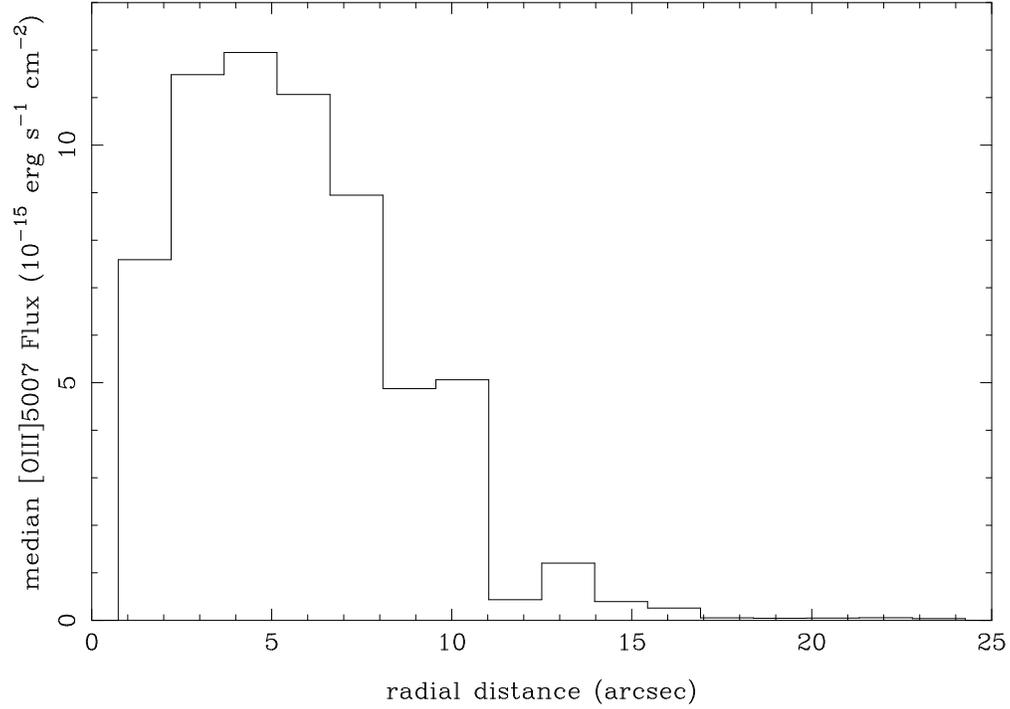}
\caption{Median value of [O{\sc~iii}] $\lambda\,5007$ on the HST image. \label{fig:OIIIradi}}
\end{figure}

\begin{figure}
\epsscale{0.8}
\plotone{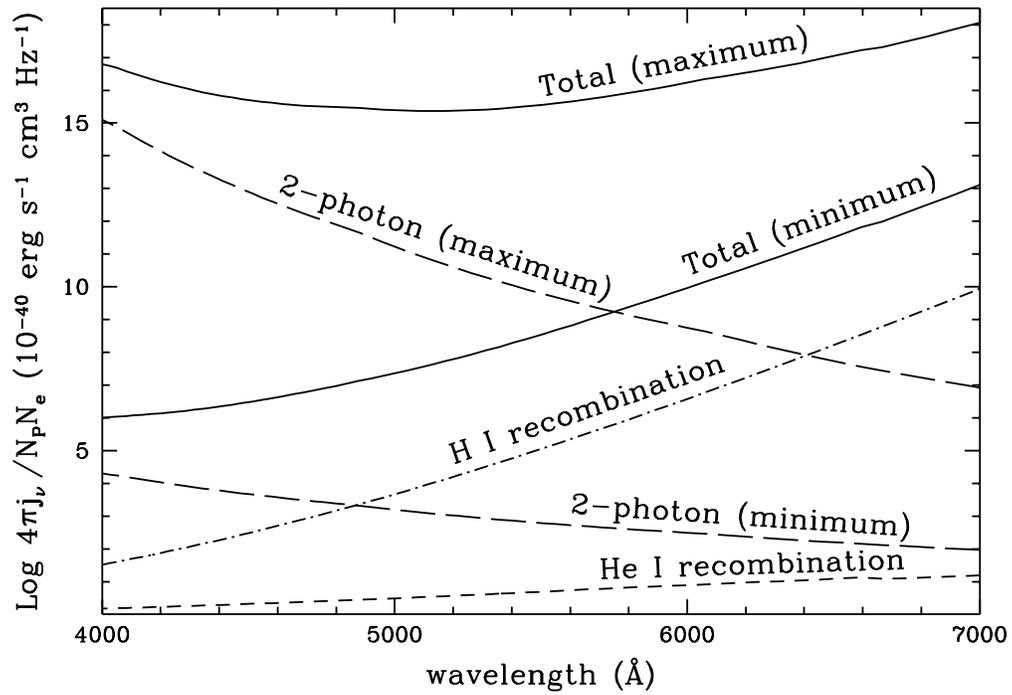}
\caption{Hydrogen and helium continua at $T_e=8000$ K. \label{fig:continuum}}
\end{figure}

The expected continuum intensity can be compared to
direct measurements of the continuum level in an optical spectrum.
We followed this procedure measuring (through the corresponding filter
responses,
as described in section~\ref{sec:linecon}) the integrated continuum flux
and
relevant line intensities in the two long-slit spectra of NGC 6543
described in Section~\ref{sec:INTdata}.
By means of this procedure, we find that the continuum contributes
58$\%$
of the total intensity in the F437N filter around [{\sc O~iii}] $\lambda\,4363$
(cf. Figure~\ref{fig:spec}), significantly more than 
expected on the basis of our previous calculations.

Three likely candidate mechanisms that would enhance the continuum
level
are X-ray emission, dust scattering, and 2-photon emission.
X-ray diffuse emission has been recently reported from high resolution spectral
images
by Chandra \citep{Cal01}; this is produced in the mass-loading shocked
region where the fast stellar wind interacts with the inner gaseous shell.
The optical tail of such emission could be the origin
of the observed extra continuum.
To test this point, we modeled the central cavity of NGC 6543 by
means of simple, spherical hot-plasma models at different temperatures,
normalized to the soft X-ray flux observed by Chandra,
and computed for each model the contribution to the F437N band.
Large uncertainties in this calculation are introduced
by the values assumed for the plasma temperature, the exact lower energy 
limit seen by Chandra (which is around 0.1 -- 0.2 keV), and the intervening absorption
by neutral gas (which we did not take into account in our estimates).
For example, a $T=1.7 \times 10^6$ K plasma, normalized to the emission in the 
(0.1 -- 10 keV)
interval, produces only 0.001\% of the observed extra-continuum in F437N,
while  a $T=5\times 10^5$ K plasma, normalized to the emission in the (0.4 -- 10 keV)
interval, produces 10\% of the observed extra-continuum in F437N.
Although these are only back-of-the-envelope computations, 
altogether they suggest that the extra continuum
cannot be explained in terms of hot plasma emission.

A different mechanism that could explain the extra continuum is dust scattering of the central
star's radiation.
Even if the external screen extinction is rather small, $C({\rm H}\beta)=0.07$
(see below), a cursory inspection of SWS+LWS archival spectra of NGC 6543 shows a
thermal
bump due to hot dust peaking at $\sim30\mu$. 
Although this mechanism is a plausible candidate 
for the origin of the extra continuum, 
the detailed modeling of the internal dust distribution in NGC 6543 
is beyond the scope of this paper. 

\begin{figure}
\epsscale{0.8}
\plotone{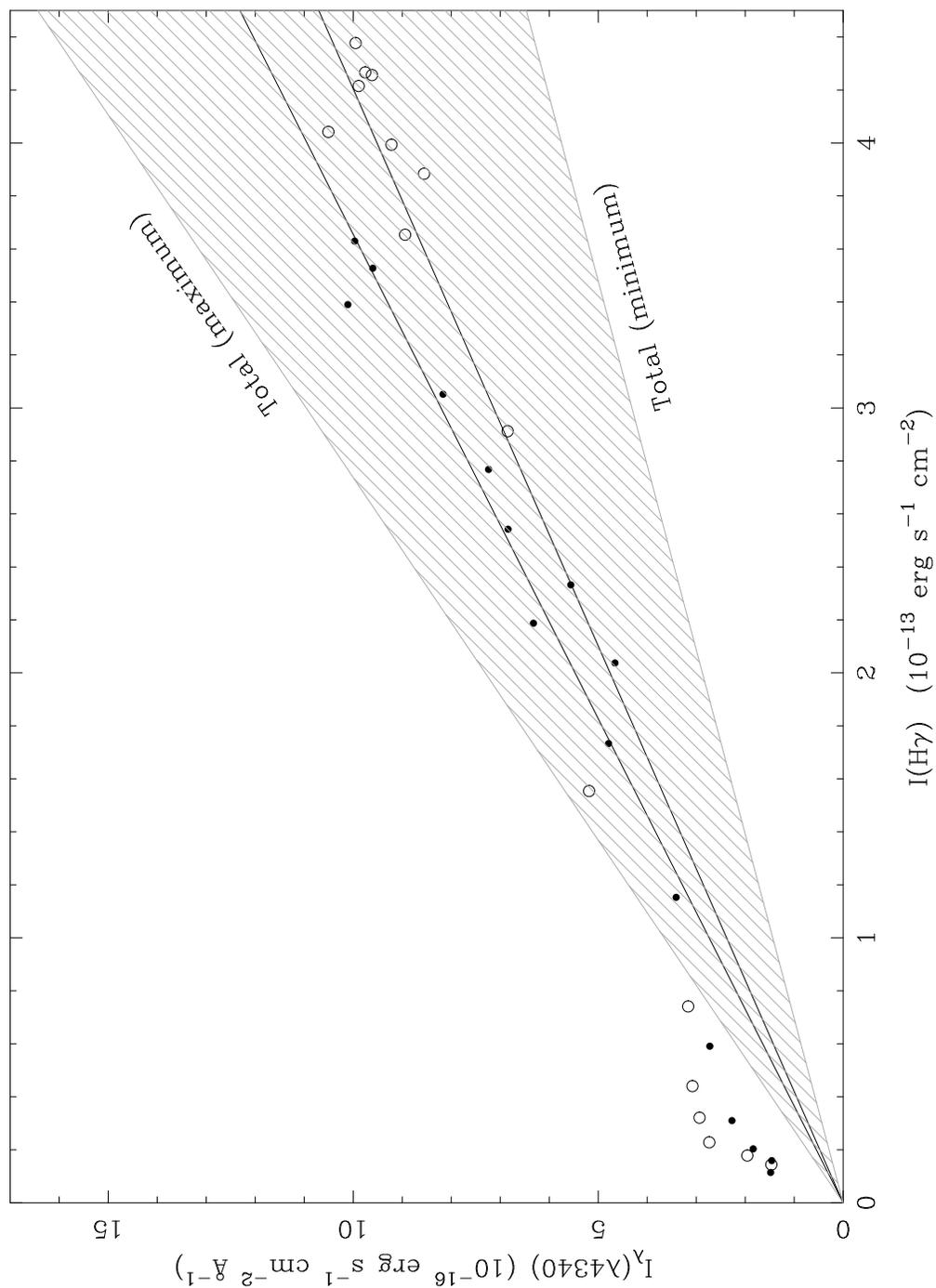}
\caption{Observed H$\gamma$ and nearby continuum fluxes
measured along the INT slit, as a function of the distance
from the central star. The shaded region corresponds to the region
predicted by recombination theory.
Dots and open circles represent the two directions along the slit,
and the two solid lines the corresponding best-fit linear relations.
\label{fig:contHgamma}}
\end{figure}

A third possibility is 2-photon emission with enhanced efficiency.
As discussed in Section~\ref{sec:cont_theo}, 
in our estimate of the 2-photon emissivity we made use of a simplified
treatment \citep[][]{O89,BM70},
which neglects Ly$\alpha$ conversion. 
The efficiency of Ly$\alpha$ conversion to produce
2-photon emission varies between
0 in optically thin or dusty nebulae (all the Ly$\alpha$ photons either escape 
or are destroyed by dust) and a maximum (all the Ly$\alpha$ photons
are eventually converted to 2-photon emission). 
The detailed calculation of the actual efficiency of this process is virtually
impossible, since it would require knowledge of the local escape probabilities
and the dust structure of the nebula, and the implementation of this information
in a 3-D computation. 
The only feasible alternative is to bracket this process
by computing the 2-photon emission in
the extreme case of maximum efficiency.
Following the formalism of \citet{BM70},
it is found that the 2-photon emission at $\lambda\,4363$ can be enhanced
by Ly$\alpha$ destruction by up to a factor of 3.5,
rising the total continuum by up to a factor of 2.5.
Figure~\ref{fig:contHgamma} compares the observational data to the continuum range
predicted by recombination theory,
as a function of the H$\gamma$ intensity. The data are
compatible with a Ly$\alpha$-conversion enhancement of the 2-photon continuum,
and suggests that it is not necessary to invoke other processes to explain
the measured continuum level
(the only discrepant points are some of those with
the lowest intensity values, i.e. with the largest errors).
The figure also shows that the relation 
between H$\gamma$ and the measured continuum is not linear,
and that there are large fluctuations around the best-fit straight line,
as expected if the continuum in excess of the minimum level 
arises from Ly$\alpha$ conversion:
stated otherwise, the Ly$\alpha$ conversion process acts with different
efficiency across the nebula.
This figure demonstrates
that it is not possible to rely on theoretical calculations alone
to correct the total flux in each image for the continuum contribution:
first because we would ignore, a priori, the average
efficiency of the Ly$\alpha$ conversion process
(the slope on the H$\gamma$-continuum relation,
which is provided by the slit spectrum);
second because, even with this piece of information available,
we would still ignore the large fluctuations of the continuum around
such average level.
These fluctations naturally arise as a consequence
of the local character of the 2-photon enhancement process,
and, since theory alone provides no way of predicting
where and how large they are across the image,
they would translate directly into an uncertainty in the measured
$\lambda\, 4363$ flux. 
Furthermore, this uncertainty would be relatively much larger
for $\lambda\, 4363$ than it is for the continuum, 
since the relative contribution of the line in the F437N filter
is less than one half that of the continuum,
so that a typical uncertainty of, e.g., 20\% in the continuum would
translate into an uncertainty of 45\% in the line.
This would be much larger than the typical uncertainty 
that can be read directly from Figure~\ref{fig:spec},
and it would not avoid the need for a spectroscopic measurement;
so this strategy to determine $I(\lambda\, 4363)$ across the image
is definitely not viable.

Note also that this conclusion holds even under the hypothesis
in which the 2-photon continuum is at its minimum level (the only case
that would allow to predict the continuum produced by recombination
by means of a recombination line)
and that the additional continuum flux comes instead from a different source:
as a matter of fact, the continuum is not proportional to H$\gamma$,
hence it cannot be predicted theoretically, 
no matter which process is responsible for the departure
from direct proportionality.
An important consequence is that no temperature maps can be obtained
using this kind of data, unless very specific circumstances occur:
one would be that $\lambda\,4363$ dominate the light in the filter,
so that only a rough correction for the continuum would suffice;
a second one would be the availability of tunable filters
to measure directly the continuum level near $\lambda\,4363$,
bypassing the need for theoretical predictions.
The lesson to be learned from these computations
is that precision photometry of weak lines cannot
be done with the HST/WFPC2 in the absence of data
specifically designed to measure the continuum.
As a matter of fact, an additional uncertainty
is introduced even by the estimation of the contribution 
of the various lines inside each filter,
unless a reliable calibration is available.
Both these problems and possible solutions to them
- continuum and line subtraction -
have been addressed by \citet{ODD99} and \citet{ODPP03};
their arguments will be described in Section~\ref{sec:cont_F437N}.
To proceed with the discussion, we will therefore assume 
that the $\lambda\,4363$ percentage within the filter
F437N is the one provided by the bottom panel of
Figure~\ref{fig:spec}, i.e.
$I(\lambda\,4363)$/$I_{tot}($F437N$) = 0.26 \pm 0.03$.
It must be emphasized, however, that in the absence of slit spectra
or sophisticated theoretical tools to estimate a priori the
continuum intensity, such uncertainty would be much larger,
amounting to the difference between the minimum and maximum
continuum value admitted by recombination theory.

In their paper on NGC 6543,
\citet{Hal01} pointed out
inconsistencies between the theoretical estimation 
and the continuum level derived from observations
(in their case, ground-based echelle spectra).
It is not clear, however, whether such inconsistencies can
be tracked down to what we point out in this work, 
as their procedure differs from ours
in at least two  relevant aspects:
first, 
the theoretical continuum flux computed by \citet{Hal01}
includes only the contributions by H{\sc~i} recombination and bremsstrahlung.
However, Fig. 1 in \citet{BM70} shows
that, at $T_e=10000$ K, 
the continuum is dominated by 2-photon continuum at $\lambda\,4363$, 
and this process
is only slightly less important than H{\sc~i} recombination
at $\lambda\,5007$.
At $T_e=8000$ K the 2-photon contribution is even more important.
The 2-photon continuum must therefore be included even in a rough estimation
of the theoretical continuum flux.
Second, 
\citet{Hal01} adopt smaller values than we did for the width of the filter:
25.2 \AA, 26.9 \AA, and 21.9 \AA \
for F437N, F502N, and F656N respectively,
instead of 31.829 \AA, 35.781 \AA, and 28.336 \AA.
Apparently, these values correspond to
the effective width of the bandpass $\delta \lambda$ 
defined by \citet{Bal01} (pag. 149 and Appendix 1; 
see also Section~\ref{sec:width} above);
however, we might be misunderstanding their procedure, 
since this point is not explicitly  stated in their paper, and
their H$\alpha$ width slightly differs from the one given by \citet{Bal01}.
But if we are understanding their procedure,
then the difference in the adopted filter widths propagates directly into
differences in the estimated continuum flux. This introduces
a large bias in all the derived fluxes.
Furthermore, since the ratios of different RECTW values are not
equal to the ratios of different $\delta \lambda$s,
the intensity ratios of different images are correspondingly affected
(the way it affects the derived $\lambda\,4363$ is more subtle,
because the continuum contribution to be subtracted 
from the F437N flux is directly proportional
to the adopted bandwidth).

It seems probable that \citet{LHB97} also relied on a theoretical 
computation to estimate the theoretical continuum in the F437N image,
since the number they quote, $F_c($F437N$) = 0.015 \times F($H$\beta$), 
is comparable (actually,
smaller) than our theoretical computation. However, their work is just a 
short conference contribution, 
and they do not specify how such estimate is obtained.

A similar procedure was also followed by \citet{Ral02} 
in their analysis of the planetary nebula NGC 7009.
These authors obtain a temperature map of 
NGC 7009 by means of WFPC2 images taken 
in the same filters considered by us.
To estimate the line contribution inside the WFPC2
filters, they make use of HST/STIS data, modified to take into account
the transmittance  of the corresponding WFPC2 filters.
Following this procedure,
they find that the continua measured from STIS are higher
than expected at the temperature of the nebula.
They interpret such discrepancies as an artefact due
to insufficient signal-to-noise in the STIS data, 
and therefore adopt in their analysis a continuum
flux from theoretical computations;
it is possible, however, that the problem they found 
with the continuum in NGC 7009 is analogous to what is found
in NGC 6543, namely 2-photon continuum enhancement.
In any case, the uncertainty on the continuum
is less crucial in their work than it is in the
case of NGC~6543, because $\lambda\,4363$ is relatively more
intense in NGC 7009:
they measure in F437N a $\lambda\,4363$/continuum ratio 
equal to 3.7, i.e. almost one order of magnitude higher
than the one observed in NGC~6543.

Summarizing this section on continuum, 
it is important to stress that there is no safe
way to predict accurately the $\lambda\,4363$ contribution 
to the total flux in the filter F437N, 
unless this line is very bright.
Because of this difficulty,
special care should be given to this point, and the
relevant assumptions should always be made explicit.

\subsection{Reddening correction}

\begin{figure}
\plotone{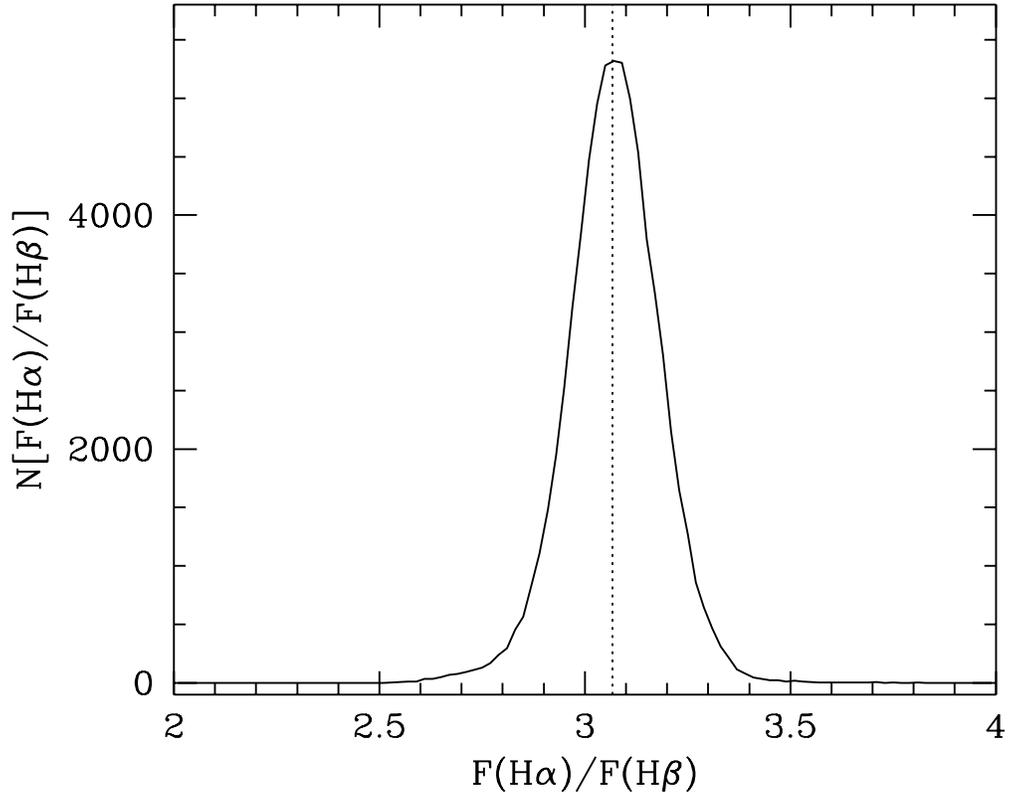}
\caption{Pixel-by-pixel $F$(H$\alpha$)/$F$(H$\beta$) histogram. \label{fig:HaoverHb}}
\end{figure}

The mean value of the H$\alpha$/H$\beta$ flux ratio across the HST image 
of NGC 6543,
with the [N{\sc~ii}] and the continuum contributions subtracted, 
is H$\alpha$/H$\beta$=3.06$\pm$0.12 (cf. Figure~\ref{fig:HaoverHb}). 
Adopting the extinction law by \citet{CCM89} with 
$R_V$=3.0, 
and fitting to the theoretical Balmer decrement at $T_e=8000$ K
\citep{SH95},
a reddening coefficient 
$C({\rm H}\beta)=0.07\pm0.06$ is obtained.

The three $C({\rm H}\beta)$ values compiled by \citet{Hal01},
$C({\rm H}\beta)$=0.18, 0.20 and 0.30,
are significantly higher than our result. 
We ignore the origin of this discrepancy, 
since none of the papers referenced by \citet{Hal01}
lists the observed H$\alpha$ fluxes, nor do they provide details
on the assumptions underlying the computation of $C({\rm H}\beta)$.
On the other hand, our adopted $C({\rm H}\beta)$ agrees
with the value determined by other authors.  
For example, \citet{K83} found $C({\rm H}\beta)$=0.10,
and the average value from the literature compilation by \citet{KSB97}
is $\langle C({\rm H}\beta)\rangle=0.09$.

\section{Density and temperature in the O$^{++}$ zone\label{sec:NT_IR}}

The line intensities obtained by the measured fluxes
after dereddening and empirically correcting for continuum contribution, are:

\begin{equation}
I(\lambda\,4363)=(3.95 \pm 0.84)\times10^{-12} \ {\rm erg \,sec^{-1}\, cm^{-2}},
\end{equation}

\noindent and

\begin{equation}
I(\lambda\,5007)=(1.93 \pm 0.29)\times10^{-9} \ {\rm erg \,sec^{-1}\, cm^{-2}},
\end{equation}

The quoted uncertainties have been estimated as follows:
the uncertainty on $C($H$\beta)$, 0.06 dex, 
translates into uncertainties of 0.07 dex and 0.06 dex 
for $\lambda\,4363$ and $\lambda\,5007$ respectively;
the continuum subtraction adds another 0.05 dex to the 
uncertainty on $I(\lambda\,4363)$ (Section~\ref{sec:cont_obs});
finally, 
further uncertainties should be added 
to account for the shift in wavelength 
of the bandpass with temperature,
and the photometric errors
\citep[cf.][Table 8.11]{Bal01};
however, this is a very complex issue,
as it will be shown in Section~\ref{sec:HSTcalib}, 
and this source of uncertainty will not be considered for the moment.

These values yield
$\lambda\,5007$/$\lambda\,4363$ = 490$\pm$130,
and $T_e($O$^{++})_{opt}=7500\pm 450$ K,
a temperature lower than previously published values. 
For example, \citet{Hal01} find $T_e= 8000 \sim 8300$ K,
\citet{KLP96} find $T_e= 7950 \pm 100$ K, \citet{PTPL95} find $T_e= 8334$ K,
and DLW adopt $\lambda\,5007$/$\lambda\,4363$ = 363 from \citet{BDO74},
which yields, using the same atomic data as us, $T_e($O$^{++})_{opt}$=7900 K.

\begin{figure}
\plotone{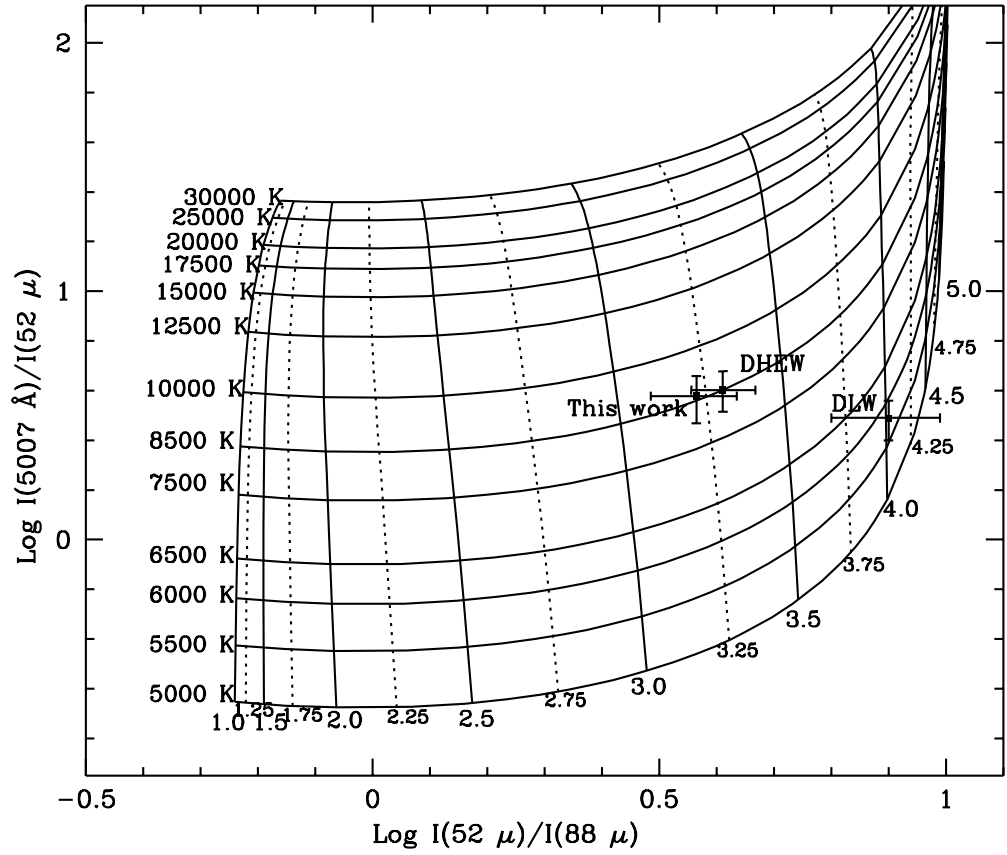}
\caption{[O{\sc~iii}] infrared-line diagnostic diagram. \label{fig:OIII}}
\end{figure}

Combining the infrared intensities measured on ISO spectra
with the [{\sc O~iii}] $\lambda\,5007$ intensity
derived from the HST data, it is possible to simultaneously derive
the temperature and density in the O$^{++}$ zone of NGC 6543 
by means of the infrared-line diagnostic described by DLW.
The diagnostic diagram is shown in Figure \ref{fig:OIII}, where the point
representing the data has been plotted. 
The data points by DLW and DHEW are also plotted.
The values obtained are $N_e = 1650^{+550}_{-400}$ cm$^{-3}$, $T_e = 8600 \pm 500$ K,
while the corresponding values quoted by DLW and DHEW are 
$N_e$= $10000^{+\infty}_{-6000}$ cm$^{-3}$, $T_e = 5800 \pm 300$ K
and $N_e$= $2000^{+500}_{-400}$ cm$^{-3}$, $T_e = 8500 \pm 500$ K
respectively.

Figure \ref{fig:OIII} and Table~\ref{tab:comparison}
show that the point by DLW and the one we obtain here
disagree by much more than $1\sigma$, both in $T_e$ and $N_e$. 
The three lines are discrepant, particularly $52\,\mu$ and $\lambda\,5007$
(however, the discrepancy in $\lambda\,5007$ depends solely on the adopted
$C($H$\beta)$, since the difference between the measured fluxes is less than 1\%).
These differences go in the same sense, 
so that the $\lambda\,5007/52\mu$ ratio is not very different in the two cases:
however, the upward bending of the isotemperature contours in the 
high-density region of the diagram enhances
the difference between the corresponding temperatures.
On the other hand, our results are in excellent agreement with those by DHEW,
though unfortunately the fluxes of individual lines cannot be compared
since they were not published. 

\subsection{The comparison between $T_e($O$^{++})_{opt}$ and $T_e($O$^{++})_{IR}$}

In the presence of temperature variations across a low-density, homogeneous nebula, 
it is well known that
$T_e($O$^{++})_{opt}$ preferentially weighs regions of high $T_e$, 
while $T_e($O$^{++})_{IR}$ preferentially weighs regions of lower $T_e$. 
In such a case $T_e($O$^{++})_{opt} \ge T_e($O$^{++})_{IR}$,
with the equality holding in the constant-temperature case.
When density fluctuations are also present, this basic scenario complicates, 
because the temperature is lower in the densest zones, 
and because collisional de-excitations of the infrared lines may play a role.
Roughly, we expect higher $T_e($O$^{++})_{IR}$ values
due to the collisional suppression of the 52 $\mu$ line;
as a result, [$T_e($O$^{++})_{opt}$ -- $T_e($O$^{++})_{IR}$] may
become negative.
Since density fluctuations are certainly present in the nebula, as revealed by the 
filaments and thin shells seen in the H$\alpha$ images of the nebula,
the value we determined for $T_e($O$^{++})_{IR}$ is totally plausible.

Summing up, our $T_e($O$^{++})_{opt}$ does only marginally 
agree with the published values;
$T_e($O$^{++})_{IR}$ agrees with the most recent of the only two determinations
we know of; and $T_e($O$^{++})_{IR}$ compares well with the expected value
of $T_e($O$^{++})_{opt}$ (higher than ours by $\sim$ 400 -- 800 K), provided
there are high density zones in the nebula, 
which depress the flux in $52\,\mu$.
In the following, we will list and discuss
the uncertainties affecting our results.
In this discussion we will consider several uncertainties
affecting all the data,
although,
in the light of the preceding discussion and the previous comparison to published data 
(see Section~\ref{sec:NT_IR}),
it seems probable that the problem lies with the adopted WFPC2 
$\lambda\,4363$ flux,
and that our ISO infrared data and the WFPC2 $\lambda\,5007$ data are 
reasonably accurate.

\section{Possible sources of error}

\subsection{Continuum subtraction in F437N\label{sec:cont_F437N}}

The long slit on which the continuum level was measured samples
only one fraction of the whole nebula,
and we cannot guarantee that such fraction is indeed
representative of the rest.
It seems not probable, however, that the continuum
varies so dramatically in the area outside the slit,
so as to alter sensibly our conclusion.
A further possibility is a bias in the continuum level
of the slit spectra.
A bias in the continuum level
could be caused by at least two different circumstances:
an inaccurate correction for light dispersion inside
the spectrograph, 
and scattered light from the central star within the nebula. 
The first effect could be important in view of the fact
that the flux in the continuum outweighs the flux in $\lambda\,4363$
in the filter, 
so that a small bias in the continuum would be amplified
in the derived $\lambda\,4363$ value.
The second effect, which would mimic a more intense nebular
continuum, 
would affect equally the slit spectra and the HST images, so it
cannot be invoked as a cause for the low $T_e($O$^{++})_{opt}$ found,
and will not be further discussed.

\citet{ODD99} stress the importance of an accurate calibration
of the WFPC2 filters
in order to make a quantitative use of the HST emission-line images.
One of the problems they discuss related to calibration
is that the filter characteristics described in the WFPC2 Handbook
have been determined in parallel light, 
whereas the incident beam inside the WFPC2 is convergent.
The consequence is that the filters {\sl as used by the WFPC2} have shorter
peak transmissions and wider FWHMs.
The use of the pre-launch filters' characteristics provided with the HST documentation
can therefore introduce a bias, and they recommend instead
the use of calibration relations based on the instrument ``as used''.
A specific problem they address is the continuum subtraction;
the strategy they suggest to determine the continuum level
is to use an image taken in F547M, a filter that includes no 
prominent emission lines, together with an assumption
on the color of the continuum.
Unfortunately, {NGC~6543} has not been observed
in this filter, so this method is not applicable in our case.

A different strategy to subtract the continuum
from the F437N filter has been proposed by \citet{ODPP03} 
for the case of low-excitation objects.
The basic idea is to take an image in the F469N filter:
the constraint on the excitation degree of the object
ensures that the contribution of the
He{\sc~ii} line $\lambda\,4686$ is negligible,
and all the signal in the image can be interpreted
in terms of continuum.
This procedure allows to reduce the uncertainty related
to the assumption on the color of the continuum,
since the distance between the wavelengths
of F437N and F469N is quite small.
Unfortunately, although the He$^{++}$/H${^+}$ ratio in NGC 6543
is very small and make this object suitable to apply 
the method by \citet{ODPP03}, no F469N image of this object has been obtained
(maybe it is precisely {\sl because} the He$^{++}$/H${^+}$ ratio in NGC 6543
is so small, that an observation of the nebula with this
filter was judged uninteresting by past observers).

\subsubsection{Other issues related to the HST/WFPC2 calibration\label{sec:HSTcalib}}

There are further possible explanations 
for our anomalous $T_e($O$^{++})_{opt}$ result
related to the WFPC2 calibration.
We will give a broad overview of them,
without estimating them quantitatively.

One problem is the accuracy in the determination
of the contribution of the various lines in each filter.
Due to the nature of the calibration procedure
mentioned in the previous section, 
together with the natural changes suffered by any filter
with time, 
the specifications reported in the WFPC2 Handbook do
not necessarily give an accurate description of
the actual filters' transmission curves.
As a result, the decontamination of the F656N flux
from the [N{\sc~ii}] $\lambda\,6584$ contribution
could be slightly biased;
and the decontamination of the F437N flux 
from the H$\gamma$ contribution could be even more
biased.
Unfortunately, the relations provided by \citet{ODD99}
to deal with this problem are not applicable,
because they make use of the flux in the F547M filter,
which has not been used to observe NGC 6543.

There are also other potential problems that
affect only faint images with very low DNs.
The composite F437N image used in the study has
a mean DN of 2.9, corresponding on average to ${\rm DN} \sim 1.5$
in each of the two exposures 
(the DN cannot be measured individually on these two images,
since it is spuriously increased by cosmic rays, which are removed
in the composite image);
since these DNs are extremely low,
the effects mentioned in the following are possibly a 
large source of uncertainty in the  F437N flux.
On the other hand, the images taken with the other filters
have  DNs of the order of hundreds,
and their calibration is much more accurate.

Further relevant sources of uncertainty, which affect 
all CCDs at low DNs,
are the read noise and the loss of information 
implied by the digitization of the signal.
Another important source of uncertainty in our specific case
is the progressive alteration
in the performance of the WFPC2 CCDs:
in orbit, the CCDs are subject to radiation damage
\citep{Bal01}, which alters their long term properties
leading to an increase in dark counts and a decrease
in the charge transfer efficiency (CTE).
Dark counts increase because of hot pixels,
which are routinely identified and can, in principle, be corrected.  
Conversely, the decrease in CTE is difficult to estimate, 
since the existing calibrations \citep{WHC99} are based on measurements of
the CTE loss for a bright point source (${\rm DN} \greatsim 60$ for gain = 7)
observed against a faint background (${\rm DN}_{bg} \lesssim 1$ for gain = 7); 
therefore, it is not clear whether the correction formulae provided
by \citet{WHC99} can be equally applied to the case
of a diffuse, nebular object, which can be seen as
a moderately faint object seen against a moderately faint background
(in our case, DN = ${\rm DN}_{bg} \sim 1.5$ for gain = 7).
On the other hand, 
Figure 4.14 in \citet{Bal01}, based on recent observations,
includes a few more points for brighter backgrounds, in addition
to those on which the calibration by \citet{WHC99} was based.
The behavior of the recently added points
suggests that the CTE does not depend on the DN
for backgrounds of order $\greatsim$ 1.4 DN
(equivalent to a DN = 2.8 at a gain of 7).
Provided the CTE behavior is smooth with respect to time,
we can roughly estimate that
the CTE loss for the F437N filter was $\lesssim$~10\%
at the time our images were taken.
 
\subsection{ISO/LWS calibration}

As explained earlier, the uncertainty quoted on our
adopted infrared line fluxes only represents the dispersion
in our data sample. Additional uncertainties might come
from the calibration itself. 
According to \citet{Sal98}, the uncertainty in the value of the flux
for a line is given by the sum (in quadrature) of three terms:
one of them is the dispersion in the measured line flux,
and the other two are related to the detectors' responsivity
calibration. Although these terms depend on the specific observation considered,
typical values are $\sim 0.05$ and $0.01$.
Adding them to the dispersion of our data points, 
we obtain an overall uncertainty of roughly 13\% for either detector.
It is possible, however, that this value still underestimates
the real uncertainty. Reasons for this are the following:
\begin{itemize}
\item[1)] Both the drift of the SW2 detector with time and
the difference between the SW1 and SW2 detectors are 
of the order of 10\% of the measured flux.
This uncertainty is not included in the previous estimation,
and should be added to the overall error budget.

\item[2)] The spectrograph has been calibrated assuming that the 
source is point-like and centered on the optical axis. 
Since our object violates at least the first of these assumptions,
a correction to the flux should be done.
As we stated before, such correction is in principle very small 
(of order $\sim$ 1\%). 
However, the beam profile of the instrument is not completely understood,
being, e.g., much narrower than predicted by optical theory 
for reasons still unknown \citep{Sal98,Gry01};
so we cannot completely exclude that the extension of the source do play a role
in our analysis.

\item[3)] Most density determinations of NGC 6543 are well in excess
of our value $N_e=1650$ cm$^{-3}$; for example, DLW found
$N_e$[O{\sc~iii}] = 10000 cm$^{-3}$ and $N_e$[O{\sc~ii}] = 4000 cm$^{-3}$,
and  \citet{Lal01} found $N_e$[Cl{\sc~iii}] = 4800 cm$^{-3}$ and 
$N_e$[Ar{\sc~iv}] = 3900 cm$^{-3}$
(but, on the other hand, DHEW found  $N_e$[O{\sc~iii}] = 2000 cm$^{-3}$).
Higher density values would be obtained by increasing the flux of $52\,\mu$
or decreasing the flux in $88\,\mu$, leading in both cases
to lower $T_e($O$^{++})_{IR}$.
\end{itemize}

As an illustrative example of the uncertainties affecting the ISO/LWS data,
we repeated the computations adopting the mean $52\,\mu$ flux 
of the SW1 detector instead of that of the SW2 detector.
This different choice changes $\lambda\,5007/52\,\mu$ by -0.05 dex and 
 $52\,\mu/88\,\mu$ by +0.05 dex, yielding $T_e = 8100$ K and $N_e=2000$ cm$^{-3}$.

\subsection{Reddening correction}

As stated earlier in this paper, we derive a $C($H$\beta)$ value much
lower than some of the values found in the literature. 
However, we see no way of increasing our result.
Furthermore, even if this were the case, it would actually 
imply just a moderate improvement in the $T_e($O$^{++})_{opt}$ result.
To illustrate quantitatively this point, we arbitrarily increased $C($H$\beta)$ from 0.07 to 0.30,
the largest of the values compiled by \citet{Hal01}. 
Given the differential extinction affecting $\lambda\,4363$ and $\lambda\,5007$,
this change translates into a moderate decrease in the intrinsic 
$\lambda\,5007/\lambda\,4363$ ratio,
implying an increase of less than $200$ K.
On the other hand, such increase in $C($H$\beta)$ would enormously affect
the $\lambda\,5007/52\,\mu$ ratio, yielding an increase of as much as $1600$ K.

\section{Summary and conclusions}

This work was motivated by the desire to determine
self-consistently  the physical conditions in the bright core of NGC 6543.
We did this
by means of two different methods, both based on oxygen line ratios
measured across the whole bright halo of the nebula.
But, as the work progressed, it became evident
that the archival data we used were not optimized for this particular task. 

The first method used is the standard nebular-to-auroral temperature diagnostic,
and the second is the diagnostic diagram based on
infrared lines developed by \cite{D83} and DLW.
The $N_e$ and $T_e$ values derived by means of the diagnostic diagram
are not compatible with those by DLW,
but are in very good agreement to the more recent result by DHEW.
On the other hand,
the nebular-to-auroral temperature we derive is somewhat lower than 
the values published in the literature. 
We discuss some possible causes for these results,
related to the accuracy of HST, ISO, and ground-based data.

As a sideproduct of this work, we find that the continuum level
differs from the one expected 
at $T_e \sim T_e($O$^{++})_{opt}$ when Ly$\alpha$ conversion is
neglected.
We investigate the possibility 
that the extra continuum could arise from a high-temperature zone, possibly a 
shocked region, as suggested by the presence of X rays;
however, we could not reproduce the required extra continuum by
means of simple plasma models of the X-ray emitting bubble.
An alternative explanation we propose is that it is generated by
dust scattering of the central star's light.
A third possibility, which is the most plausible,
is that the additional continuum is enhanced 2-photon emission,
originated by conversion of scattered Ly$\alpha$ photons.

In neither case is it possible to work out a theoretical prescription
to compute accurately the continuum intensity that would 
eliminate the necessity of relying on spectroscopic information.
Furthermore, either of these mechanisms would affect equally the slit
spectra and the image, so neither would explain the marginally low
temperature value that we deduce by means of the HST data.
As a result, the most important source of uncertainty in the determination of the 
optical temperature is the continuum subtraction in the $\lambda\,4363$ image;
a specific conclusion we draw 
is that it is not possible with these data to obtain an accurate bidimensional
temperature map of the nebula.
More generally, we review several possible causes of error that
could be introducing a bias in our work.
The calibration uncertainty of WFPC2 at very low DNs introduces perhaps an
additional uncertainty.
The HST fluxes at other wavebands and the ISO data are probably more robust,
but the overall uncertainty could still be somewhat higher than 10$\%$. 
In the case of HST data, a fraction of this uncertainty might arise 
from the contribution of different lines to the observed fluxes; 
in the case of ISO data, the overall accuracy is probably of order $\sim$ 13\%.

Although NGC 6543 has been previously studied by many authors, 
most measurements have been made with small apertures which sample
only a small fraction of the nebula. One disadvantage of this approach is
that it limits the possibility to compare among different results.
On the contrary, the data we use sample the whole region,
allowing us to do a self-consistent analysis which eliminates the need 
of assumptions about the ionization structure.
Furthermore, the use of lines of the same ion allows 
a self-consistent determination of $N_e$ and $T_e$,
independent of assumptions on the ionization structure.
Unfortunately, the available archival data were not specifically 
designed for this analysis,
and contain sources of uncertainties that are, in many cases,
quite difficult to estimate.
We think therefore that our results on the physical conditions
in NGC 6543 are somewhat biased,
and we definitely exclude the possibility of using
these data to derive sensible temperature maps
unless specifically designed data are obtained that provide a safe way of
measuring the continuum and line contribution inside each filter.
Our results are, however, largely informative for what concerns 
delicate technical issues
that must be taken into account when studies of this kind are carried out.
In particular, we draw the attention, throughout this paper, 
to complex aspects of the use of data from large archives, 
in the hope that the authors of future similar works will be motivated 
to provide an explicit description
of the crucial assumptions underlying their analyses.

\begin{acknowledgements}
We thank Jes\'us Ma\'\i z-Apell\'aniz 
for helping us through the jungle of WFPC2 data.
We are also very grateful to Gra\.zyna Stasi\'nska for
an early reading of this paper and several useful suggestions,
and Robert O'Dell and Antonio Mampaso for insightful comments.
It is also a pleasure to acknowledge Chris Lloyd and the ISO helpdesk
for their keen answers to our questions.  
Support from the HST helpdesk is also acknowledged.
We thank Luis Cuesta, David Axon and Andrew Robinson, who contributed
to obtain the ground based spectroscopic data.
Two anonymous referees provided us with several useful suggestions
and criticism, which were essential to improve the paper.
VL is supported by a Marie Curie Fellowship
of the European Community programme 
{\sl ``Improving Human Research Potential and the Socio-economic Knowledge Base''} 
under contract number HPMF-CT-2000-00949.
This work has been partially financed by DGICYT grants
PB98-0521 and AYA-2001-3939-C03-01.
This research has made use of NASA's Astrophysics Data System Bibliographic Services,
and it is really difficult to figure out how it was possible
to work before the ADS era.
This work is based on observations made with the NASA/ESA Hubble 
Space Telescope, 
obtained from the data archive at the Space Telescope Science Institute. 
STScI is operated by the association of Universities for Research 
in Astronomy, Inc. under the NASA contract  NAS 5-26555. 
The Isaac Newton Telescope is operated on the island of La Palma 
by the Isaac Newton Group of Telescopes in the Spanish Observatorio del 
Roque de Los Muchachos of the Instituto de Astrof\'\i sica de Canarias.

\end{acknowledgements}

\clearpage

\begin{deluxetable}{llll}
\tablecolumns{4} 
\tablewidth{0pc} 
\tablecaption{ISO LWS archive spectra.\label{tab:ISOjournal}}
\tablehead{ 
\colhead{Epoch} & \colhead{Date} & \colhead{TDT}  & Proposal ID \\}
\startdata 
Epoch I  & 05-Jun-1996 & 20101905 & CAL\_R196 \\ 
          & 07-Jun-1996 & 20301905 & CAL\_R203 \\ 
          & 14-Jun-1996 & 21003105 & CAL\_R210 \\ 
          & 21-Jun-1996 & 21704705 & CAL\_R217 \\ 
          & 27-Jun-1996 & 22301705 & LWS\_R223 \\
          & 04-Jul-1996 & 23001905 & CAL\_R230 \\
          & 11-Jul-1996 & 23705405 & CAL\_R237 \\
          & 18-Jul-1996 & 24404805 & CAL\_R244 \\
          & 24-Jul-1996 & 25100805 & CAL\_R251 \\
          & 28-Jul-1996 & 25500701 & DUST4\_A  \\ 
\smallskip
          & 31-Jul-1996 & 25800905 & CAL\_R258 \\
Epoch II & 02-Dec-1997 & 74801614 & CAL\_R748 \\ 
          & 09-Dec-1997 & 75500604 & CAL\_R755 \\ 
          & 16-Dec-1997 & 76201203 & CAL\_R762 \\ 
          & 23-Dec-1997 & 76902404 & CAL\_R769 \\ 
          & 30-Dec-1997 & 77602002 & CAL\_R776 \\ 
          & 06-Jan-1998 & 78300202 & CAL\_R783 \\ 
          & 13-Jan-1998 & 79000702 & CAL\_R790 \\ 
          & 20-Jan-1998 & 79700402 & CAL\_R797 \\ 
          & 27-Jan-1998 & 80401402 & CAL\_R804 \\ 
\enddata 
\end{deluxetable} 

\begin{deluxetable}{lllcccc}
\tablecolumns{7} 
\tablewidth{0pc} 
\tablecaption{ISO/LWS fluxes.\label{tab:ISOdata}}
\tablehead{ 
\colhead{} & \colhead{} & \colhead{} & \multicolumn{2}{c}{Line intensity\tablenotemark{a}} & \multicolumn{2}{c}{Continuum flux\tablenotemark{b}}\\
\cline{4-5} \cline{6-7}\\
\multicolumn{1}{c}{Line} & \multicolumn{1}{c}{Detector} & \multicolumn{1}{c}{Epoch\tablenotemark{c}} & \colhead{Median} & \colhead{$\sigma$} & \colhead{Median} & \colhead{$\sigma$} \\
}
\startdata 
$52\mu$&SW1& I    &5.58&0.50&1.66&0.29\\
       &   & II   &6.10&0.43&1.91&0.68\\
       &   & I+II &5.74&0.53&1.70&0.50\\
\cline{2-7}
       &SW2& I    &5.06&0.66&1.54&0.11\\
       &   & II   &5.15&0.52&1.49&0.12\\
       &   & I+II &5.10&0.61&1.53&0.12\\
\hline\hline
$88\mu$&SW5& I    &1.40&0.18&2.69&0.47\\
       &   & II   &1.42&0.13&2.89&0.29\\
       &   & I+II &1.41&0.16&2.79&0.41\\
\cline{2-7}
       &LW1& I    &1.34&0.14&2.92&0.19\\
       &   & II   &1.44&0.20&3.09&0.17\\
       &   & I+II &1.39&0.17&3.02&0.20\\
\enddata 
\tablenotetext{a}{In units 10$^{-10}$ erg sec$^{-1}$ cm$^{-2}$.}
\tablenotetext{b}{In units 10$^{-10}$ erg sec$^{-1}$ cm$^{-2}$ $\mu^{-1}$ for SW1 and SW2, and 10$^{-11}$ erg sec$^{-1}$ cm$^{-2}$ $\mu^{-1}$ for SW5 and LW1.}
\tablenotetext{c}{See text and Table~\ref{tab:ISOjournal}}.
\end{deluxetable}

\begin{deluxetable}{lccrcc}
\tablecolumns{4} 
\tablewidth{0pc} 
\tablecaption{HST archive images.\label{tab:HSTjournal}}
\tablehead{ 
\colhead{Line\tablenotemark{a}}  & \colhead{Filter} & \colhead{Dataset}& \colhead{Exposure }& \colhead{PHOTFLAM}& \colhead{RECTW}\\
&&&\multicolumn{1}{c}{(s)}&\multicolumn{1}{c}{(erg s$^{-1}$ cm$^{-2}$ \AA$^{-1}$)}&\multicolumn{1}{c}{(\AA)}\\
} 
\startdata 
$[$O{\sc~iii}$]\,\lambda\,4363$& F437N & U27Q0103T & 1200$\,\,$ &7.400 $\times$ 10$^{-16}$ & 31.829\\
                               &       & U27Q0104T & 1200$\,\,$ &      &       \\
H$\beta$                       & F487N & U27Q0105T &  500$\,\,$ &3.945 $\times$ 10$^{-16}$ & 33.921\\
                               &       & U27Q0106T &  900$\,\,$ &      &       \\
$[$O{\sc~iii}$]\,\lambda\,5007$& F502N & U27Q0107T &  200$\,\,$ &3.005 $\times$ 10$^{-16}$ & 35.781\\
                               &       & U27Q0108T &  600$\,\,$ &      &       \\
                               &       & U27Q0109T &  200$\,\,$ &      &       \\
                               &       & U27Q010AT &  600$\,\,$ &      &       \\
H$\alpha$                      & F656N & U27Q010FT &  200$\,\,$ &1.461 $\times$ 10$^{-16}$ & 28.336\\
                               &       & U27Q010GT &  600$\,\,$ &      &       \\
$[$N{\sc~ii}$]\,\lambda\,6584$ & F658N & U27Q010HT &  500$\,\,$ &1.060 $\times$ 10$^{-16}$ & 39.232\\
                               &       & U27Q010IT &  500$\,\,$ &      &       \\
\enddata 
\tablenotetext{a}{Most important emission line included in the passband.}
\end{deluxetable} 

\begin{deluxetable}{lrrrr}
\tablecolumns{5} 
\tablewidth{0pc} 
\tablecaption{Frequency dependence of some continuum-emission coefficients
(in 10$^{-40}$ erg sec$^{-1}$ cm$^3$ Hz$^{-1}$) at $T_e=8000$ K.\label{tab:continuum}} 
\tablehead{ 
\colhead{}    &  \multicolumn{1}{c}{$\lambda\,4363$} &  \multicolumn{1}{c}{$\lambda\,5007$} &  \multicolumn{1}{c}{$\lambda\,6563$} \\}
\startdata 
H I\tablenotemark{a}                &  2.196  & 3.687   &  8.431 \\
He I\tablenotemark{a}               &  2.544  & 4.502   &  10.12 \\
2-photon continuum, minimum\tablenotemark{b} &  3.827   &  3.193   &  2.175 \\
2-photon continuum, maximum\tablenotemark{c} &  13.429  &  11.204  &  7.632 \\
\enddata 
\tablenotetext{a}{Including recombination and bremsstrahlung.}
\tablenotetext{b}{Neglecting Ly$\alpha$-to-2-photon conversion.}
\tablenotetext{b}{Including maximal Ly$\alpha$-to-2-photon conversion.}
\end{deluxetable} 

\begin{deluxetable}{lrcccc}
\tablecolumns{4} 
\tablewidth{0pc} 
\tablecaption{Summary of fluxes and diagnostics.\label{tab:comparison}} 
\tablehead{ 
\colhead{}     &\multicolumn{2}{c}{DLW}& \multicolumn{2}{c}{This work} \\
}
\startdata 
\colhead{}     &\multicolumn{1}{c}{$F(\lambda)$} &\multicolumn{1}{c}{$I(\lambda)$}&\multicolumn{1}{c}{$F(\lambda)$}&\multicolumn{1}{c}{$I(\lambda)$} \\
$\lambda\,5007$\tablenotemark{a}      & $1.65\pm0.25$\tablenotemark{b} & $2.70\pm0.40 $ & $1.65$\tablenotemark{c}& $1.93 \pm 0.29$  \\
$\lambda\,4363$\tablenotemark{d}   & \nodata      & \nodata            & $3.28\pm0.40$  & $3.95\pm 0.84$\\
$52\mu$\tablenotemark{a}           &  & $8.79\pm0.97$  & & $5.10\pm0.61$   \\
$88\mu$\tablenotemark{a}           &  & $1.10\pm0.20$  & & $1.39\pm0.17$   \\
\hline
log $\lambda\,5007/\lambda\,4363$     & \multicolumn{2}{c}{$2.56$} & \multicolumn{2}{c}{$2.69\pm0.10$} \\
log $\lambda\,5007/52\mu$             & \multicolumn{2}{c}{$0.49^{+0.07}_{-0.09}$} & \multicolumn{2}{c}{$0.58^{+0.07}_{-0.09}$} \\
log $52\mu/88\mu$                     & \multicolumn{2}{c}{$0.90^{+0.09}_{-0.10}$} & \multicolumn{2}{c}{$0.56^{+0.07}_{-0.08}$} \\
\hline
$T_e($O$^{++})_{opt}$     & \multicolumn{2}{c}{$7900$}                 & \multicolumn{2}{c}{$7500\pm 450 $} \\
$T_e($O$^{++})_{IR}$      & \multicolumn{2}{c}{$5800 \pm 300$}         & \multicolumn{2}{c}{$8600\pm 500 $}       \\
log $N_e($O$^{++})_{IR}$ & \multicolumn{2}{c}{$4.0^{+\infty}_{-0.4}$} & \multicolumn{2}{c}{$3.22\pm{0.12}$}      \\
\enddata
\tablenotetext{a}{In units of 10$^{-9}$ erg s$^{-1}$ cm$^{-2}$.}
\tablenotetext{b}{Computed by us with the $C($H$\beta)$ quoted by DLW.}
\tablenotetext{c}{No uncertainty have been quoted on $F(\lambda\,5007)$, since our
estimated uncertainty on $F(\lambda\,5007)$ includes only the uncertainty on the reddening
correction (Section~\ref{sec:NT_IR}).}
\tablenotetext{d}{In units of 10$^{-12}$ erg s$^{-1}$ cm$^{-2}$.}
\end{deluxetable} 

\end{document}